\documentclass[journal,compsoc]{IEEEtran}

\usepackage{graphicx}
\usepackage{subcaption}
\usepackage{balance}
\usepackage{comment}
\usepackage{amsmath}
\usepackage{tabularx}
\usepackage{makecell}
\usepackage{adjustbox}

\usepackage[ruled,linesnumbered]{algorithm2e}
\SetAlFnt{\small\sffamily}
\usepackage[bookmarks=false]{hyperref}

\ifCLASSINFOpdf
\else
\fi

\usepackage{balance}

\usepackage{multirow}
\usepackage[table]{xcolor}
\begin{document}
	\title{HALLS: An Energy-Efficient Highly Adaptable Last Level STT-RAM Cache for Multicore Systems}
	

	\author{Kyle Kuan,~\IEEEmembership{Student Member,~IEEE} and 
	Tosiron~Adegbija,~\IEEEmembership{Member,~IEEE}

    \thanks{The authors are with the Department of Electrical and Computer Engineering, The University of Arizona, USA, e-mail: \{ckkuan, tosiron\}@email.arizona.edu.}
    \thanks{Copyright \copyright 2019 IEEE. Personal use of this material is permitted. However, permission to use this material for any other purposes must be obtained from the IEEE by sending an email to pubs-permissions@ieee.org.}
}

\IEEEtitleabstractindextext{
\begin{abstract}

Spin-Transfer Torque RAM (STT-RAM) is widely considered a promising alternative to SRAM in the memory hierarchy due to STT-RAM's non-volatility, low leakage power, high density, and fast read speed. The STT-RAM's small feature size is particularly desirable for the last-level cache (LLC), which typically consumes a large area of silicon die. However, long write latency and high write energy still remain challenges of implementing STT-RAMs in the CPU cache. An increasingly popular method for addressing this challenge involves trading off the non-volatility for reduced write speed and write energy by relaxing the STT-RAM's data retention time. However, in order to maximize energy saving potential, the cache configurations, including STT-RAM's retention time, must be dynamically adapted to executing applications' variable memory needs. In this paper, we propose a highly adaptable last level STT-RAM cache (HALLS) that allows the LLC configurations and retention time to be adapted to applications' runtime execution requirements. We also propose low-overhead runtime tuning algorithms to dynamically determine the best (lowest energy) cache configurations and retention times for executing applications. Compared to prior work, HALLS reduced the average energy consumption by 60.57\% in a quad-core system, while introducing marginal latency overhead.   

\end{abstract}

}

\maketitle

\section{Introduction}

Multicore architectures have become mainstream due to the growing demand of compute- and memory-intensive applications. Consequently, to bridge the processor-memory performance gap, much effort is being placed on designing more efficient memory hierarchies, especially for resource-constrained systems. Such designs typically involve provisioning the system with a larger last level cache (LLC) to enable higher computing throughput and alleviate challenges associated with limited main memory bandwidth \cite{Wulf95}. For example, the ARM Cortex A15 \cite{CortexA15area} allows implementations that feature four cores with a shared 1MB L2 cache. However, the LLC, which is typically implemented using conventional SRAM, imposes significant overheads with respect to leakage power and silicon area; these overheads could be prohibitive for resource-constrained systems. 

To address some of these challenges, non-volatile memory (NVM) technologies have emerged as a viable alternative to SRAMs for implementing LLCs \cite{Wolf10}. Among several emerging NVM technologies, the \textit{Spin-Transfer Torque RAM (STT-RAM)} is considered to be one of the most promising candidates to replace the SRAM \cite{Diao07}. STT-RAM, apart from its non-volatility, has other advantages, including high storage density, low leakage current, and compatibility with CMOS technology \cite{Apalkov13}. However, implementing caches using STT-RAMs is still challenging due to the overheads imposed by STT-RAM's long write latency and high dynamic write energy \cite{Sun11}. Prior studies have revealed that the STT-RAM consumes 6-14 times more energy per write access than the SRAM \cite{Sun09}. 

A popular approach for reducing STT-RAM's write latency and energy involves reducing the STT-RAM cell's data retention time. Smullen at al. \cite{Smullen11} showed that relaxing the retention time---the duration for which data blocks remain in memory in the absence of power---can substantially reduce both latency and energy. Furthermore, Jog et al. \cite{Jog12} observed that several benchmarks did not require data retention time of more than one second. Consequently, the STT-RAM's intrinsic retention time, which could be up to ten years, is unnecessary, and even undesirable in terms of energy and latency. However, the reduced retention time can sometimes be shorter than the cache blocks' requirements. To maintain data correctness, prior works propose the \textit{dynamic refresh scheme (DRS)} \cite{Li13,Smullen11,Sun11,Jog12}, which refreshes data blocks to prevent premature expiry. The refreshes incur additional overhead, which are especially worse in the LLC. Compared to the level one (L1) cache, the LLC typically requires longer cache block lifetime for data reuse. In addition, a larger number of cache blocks---as is the case in the LLC---increase the minimum number of refreshes required to hold cached data, which can drastically impact the scalability of the STT-RAM cache \cite{DASCA}.

Our work aims to mitigate the overheads imposed by STT-RAM's write energy, especially in resource-constrained systems. To this end, the work proposed herein is inspired by two important observations: 1) due to the variability in data block lifetimes, different applications may require  different retention times, and these requirements may change during runtime and 2) different applications, and combination of applications (i.e., workloads) may require different STT-RAM cache configurations (i.e., cache size, line size, and associativity) in order to minimize the energy consumption.

In this paper, we propose and explore a \textit{highly adaptable last level STT-RAM cache (HALLS)} as a viable alternative for reducing STT-RAM's write energy for LLC implementation. HALLS exploits the synergies of cache configurability \cite{Tosi16} and retention time adaptability. HALLS features a multi-banked cache that enables cache configuration through bank shutdown (to configure the cache size), bank concatenation (to configure the associativity), and multi-line fetch (to configure the line size). The different cache banks are provisioned with different retention times that can satisfy a variety of applications' requirements. Based on runtime profiling, data blocks are opportunistically placed in cache banks that offer the right-provisioned retention time for energy minimization, without substantially degrading the latency. 

Our contributions are summarized as follows:

\begin{description} 
\item[$\bullet$] We propose to design the STT-RAM LLC with different retention times in the different cache banks, and at finer granularity than prior work \cite{Sun11}. Furthermore, our work leverages the synergy of adapting both the cache configuration and retention time to different application requirements. 
\item[$\bullet$]We explore and evaluate simple and easy-to-implement algorithms to determine the best cache configurations and retention time for each executing application. 
\item[$\bullet$]We show, through extensive analysis, that compared to prior work in a quad-core system, HALLS can reduce the energy by an average of 60.57\%, while introducing minimal hardware overhead and a latency overhead of 1.47\%. Compared to SRAM, HALLS achieved average energy savings of 70.12\%, with a latency overhead of 5.16\%.
\end{description}

\vspace{-3pt}
\section{Background, Related Work, and Motivation}

From the system-level perspective, two popular approaches exist for mitigating STT-RAM's write latency and energy overheads in caches. The first involves removing as many unnecessary writes as possible. For example, dead write prediction (e.g., DASCA \cite{DASCA}) identifies dead writes---blocks that are written to the cache, but not reused thereafter---and bypasses cache accesses for those blocks. Wang et al. \cite{Wang14} also used dead write prediction to guide block placement in a hybrid SRAM/STT-RAM bank LLC. Flip-N-Write (FNW) \cite{FNW} uses bit-wise comparisons to detect the difference between a block to be replaced, and the new block. The replaced block's contents are then updated by flipping bits that are different, in order to minimize unnecessary bit-writes, thereby reducing the write energy consumption as well as the latency. 
Similarly, the Encoded Content-Aware cache Replacement (ECAR) scheme \cite{ECAR} features a block replacement policy that reduces the number of switching bits by replacing the block whose contents are most similar to the missed block. While these prior works do not enable runtime adaptability, we consider them to be orthogonal and complementary to the work presented herein. 

The second approach, on which we focus the related work discussion herein, involves substantially relaxing the retention time and incorporating the dynamic refresh scheme to ensure data correctness after the retention time elapses \cite{Smullen11,Sun11,Jog12,Li13,Qiu16}. 

In this section, we present a brief background and overview of related work in order to motivate our approach. For brevity, we omit details of the circuit-level design \cite{Diao07}, since that is not the focus of our work. 

\subsection{Refresh Schemes in Volatile STT-RAM Cache}
STT-RAM stores data bits using a magnetic tunnel junction (MTJ) cell \cite{Diao07,Dong08}. Prior work \cite{Smullen11} has shown that decreasing the MTJ cell's thermal stability can substantially reduce the STT-RAM's write latency and write energy. In effect, the MTJ cell only retains data for a limited time period---the retention time---beyond which the data would become unstable and lose correctness. To maintain data correctness, prior work \cite{Sun11,Jog12,Li13} proposed techniques that dynamically refresh the cache blocks after the retention time has elapsed. Sun et al. \cite{Sun11} used a global clock to track all valid blocks and refreshed the blocks as needed. Jog et al. \cite{Jog12} refreshed only the first eight most recently used (MRU) blocks. The authors used a write buffer to handle the surge of refresh requests, given STT-RAM's long write time. Other techniques used compiler-assisted techniques to optimize data object organization in order to make refreshes more efficient \cite{Li13,Qiu16}. However, compiler-oriented techniques, which are typically static, are not amenable to dynamic runtime changes in application requirements. 

\subsection{Overhead of DRS}\label{sec:DRS_overhead}

One of the key drawbacks of the dynamic refresh scheme (DRS) is the overhead accrued as a result of the refresh operations \cite{Jog12}. Typically, DRS requires a buffer that holds data blocks during the refresh operations. In \cite{Jog12}, the authors used a 121.6 KB STT-RAM buffer that had 1900 slots. Assuming a 128 KB direct mapped buffer with 16B line sizes and a 10ms retention time, this buffer can consume up to 141.425mW of leakage power. Ideally, the buffer should have high associativity and larger block sizes, thus consuming even more power. Furthermore, each refresh operation consists of four physical operations: 1) STT-RAM cache read, 2) buffer write, 3) buffer read, and 3) STT-RAM cache write. Given a 1MB L2 STT-RAM cache with 100$ms$ retention time, for example, our analysis revealed that each refresh operation would accrue about 1.311nJ in energy.

	\begin{figure}[t]
		\vspace{-15pt}
		\centering
		\includegraphics[width=0.9\linewidth]{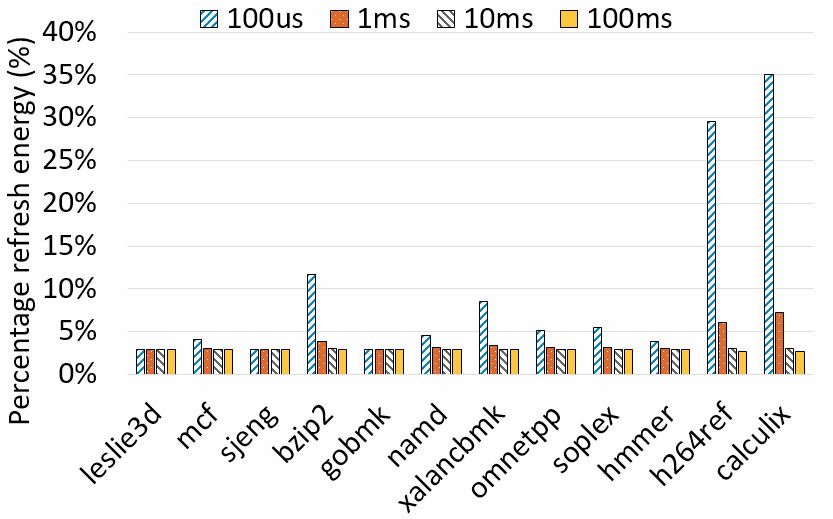}
		\vspace{-7pt}
		\caption{Refresh energy percentage for different retention times}
		\vspace{-7pt}
		\label{fig:refreshE}
        \vspace{-5pt}
	\end{figure}
	
	\begin{figure*}[t]
\centering
\begin{subfigure}[t]{.2\textwidth}
  \centering
  \includegraphics[width=.9\linewidth]{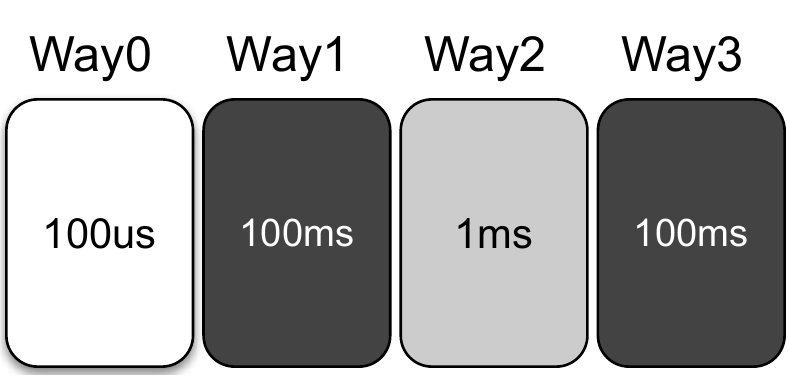}
  \caption{Workload1}
  \label{fig:Workload1}
\end{subfigure}%
~
\begin{subfigure}[t]{.2\textwidth}
  \centering
  \includegraphics[width=.9\linewidth]{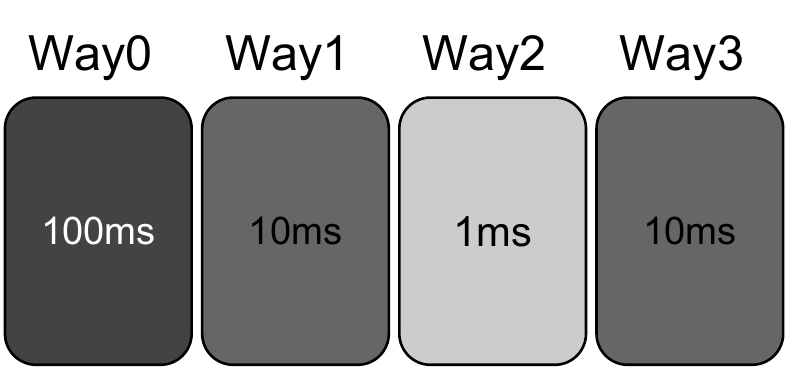}
  \caption{Workload2}
  \vspace{-7pt}
  \label{fig:Workload2}
\end{subfigure}%
~
\begin{subfigure}[t]{.2\textwidth}
  \centering
  \includegraphics[width=.9\linewidth]{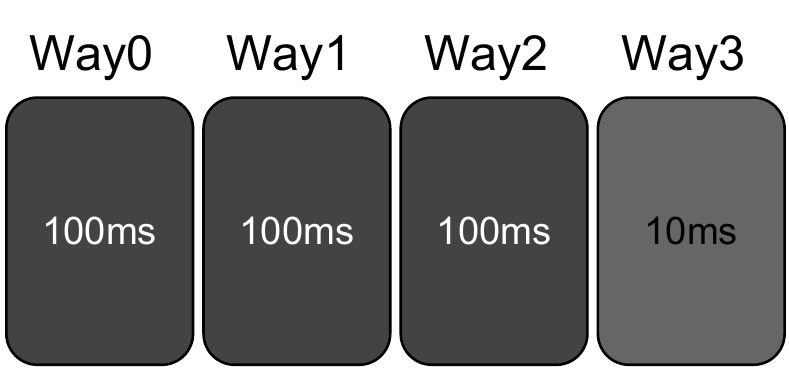}
  \caption{Workload3}
  \vspace{-7pt}
  \label{fig:Workload3}
\end{subfigure}%
~
\begin{subfigure}[t]{.2\textwidth}
  \centering
  \includegraphics[width=.9\linewidth]{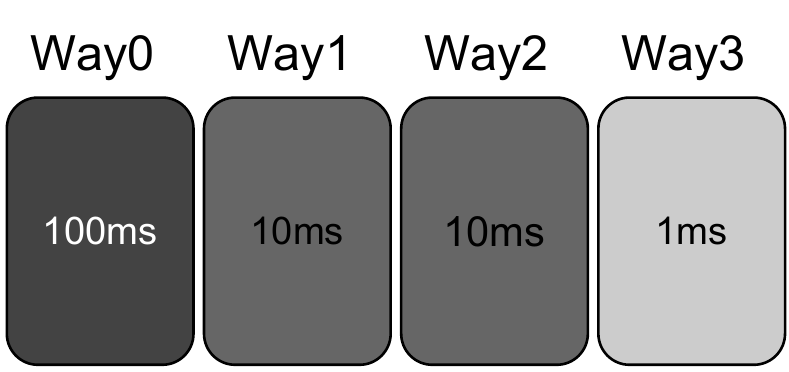}
  \caption{Workload4}
  \vspace{-7pt}
  \label{fig:Workload4}
\end{subfigure}
\vspace{-7pt}
\caption{Best energy retention time selection for different workloads in a 128KB, 64B line size, 4-way L2 cache. Each multi-programmed workload comprises of four benchmarks run in a quad-core system}
\label{fig:workload_retention}
\vspace{-10pt}
\end{figure*}

Similarly, the refresh scheme used in \cite{Sun11} accrued overheads due to the refresh buffer and its peripheral arbitration circuits. As such, leakage power was accrued during the refresh process, especially since every single L2 cache write took 3ns to 4ns. Also, even though the DASCA technique, for example, got rid of the need for refreshes by directly predicting and removing dead writes, the technique still accrued a 6.44 KB overhead for the SRAM buffer that was used to store the prediction table for a 1MB STT-RAM cache.

To further illustrate the impacts of refreshes on overall energy, we analyzed the refresh energy for different retention times while running a subset of SPEC CPU2006 benchmarks. Fig. \ref{fig:refreshE} depicts the percentage of the overall energy that comprises of refresh energy (details of our experimental setup are in Section \ref{sec:experiments}). 

We observed that for some memory-intensive benchmarks like \textit{leslie3d} and \textit{bzip2}, the total number of refreshes, and thus, refresh energy, was very low. These benchmarks had several blocks that exhibited short lifetimes and were not needed in the cache beyond the retention times. However, the refresh mechanism still constituted about 3\% of the total energy as a result of the buffer's leakage power. On the other hand, compute-intensive benchmarks like \textit{h264ref} and \textit{calculix} had blocks with longer lifetimes and higher number of refreshes, resulting in higher refresh energy. For example, \textit{calculix}'s refresh energies were 35.06\%, 7.25\%, 3.08\%, and 2.69\% of the total energy for the 100$\mu$s, 1$ms$, 10$ms$, and 100$ms$ retention times, respectively. Based on these observations, we sought to develop techniques to minimize---if possible, eliminate---the need for refreshing the cache blocks.

\subsection{Adaptable STT-RAM caches}\label{sec:priors}
Prior works \cite{Imani16,Chen12,Wang14} have discussed the benefits of resource specialization in the STT-RAM cache. Most of these works focused on leveraging a hybrid SRAM/STT-RAM cache to gain the benefits of both SRAM's low access latency and STT-RAM's low leakage power. In general, these prior works proposed data placement techniques to determine when data blocks should be placed in either SRAM or STT-RAM array. For example, Chen et al. \cite{Chen12} proposed techniques to monitor the access counts of a set and predicted the set's future usage. Similarly, Wang et al. \cite{Wang14} used execution characteristics on the CPU, such as prefetch and current program counter (PC), to determine block placement. Imani et al. \cite{Imani16} proposed a cache swap policy to intentionally place majority-zero data in STT-RAM and majority-one in SRAM, thus reducing the number of writes of '1' in order to reduce the STT-RAM's energy consumption. 

Another category of techniques for enabling STT-RAM cache adaptability focuses on profiling running applications and adapting the cache configurations based on the applications' cache requirements \cite{Tosi16,LARS}. These works focus on optimizing the L1 cache due to its impact on overall processor performance and energy consumption. Our work focuses on optimization opportunities for the shared last level caches since multicore systems are becoming increasingly ubiquitous in resource-constrained systems \cite{qcom}. However, we note that the prior techniques discussed herein are orthogonal to our work, and can be complementary to our proposed approach.

\section{Highly Adaptable Last Level STT-RAM (HALLS) Cache}
\subsection{Access Pattern and Retention Time Analysis in LLC}\label{sec:location}
Unlike the L1 cache, which is usually separated into instruction and data caches, L2 caches---we use the L2 cache to represent the LLC in this work---are usually unified. That is, L2 caches do not distinguish between instruction and data cache blocks. Since instruction blocks would not be updated or written into from CPU, they simply rely on the memory cell to hold the block's value. Thus, intuitively, instruction blocks would require longer retention times than data blocks. Data cache blocks, on the other hand, may be updated frequently by the CPU, and would require a new \textit{retention period} every time they are updated by a higher level cache (i.e., the L1 cache in our work). 

Based on these observations, we hypothesized that LLC cache blocks can be sorted into different \textit{retention time groups} depending on an executing application's code and data object behaviors. To test this hypothesis, we performed experiments using a quad-core processor featuring a shared 4-way, 128KB L2 cache with 64B blocks. We grouped ten multi-programmed workloads' data blocks into way-sized chunks of 32KB each, performed an exhaustive design space exploration of four retention times for each chunk, and then calculated the energy for each run. We empirically selected the retention times to satisfy a variety of applications' requirements, but note that these retention times may change for a different set of benchmarks. 

Fig. \ref{fig:workload_retention} illustrates the combination of retention times that achieved the lowest energy for a workload (for brevity, only four workloads are shown, but the analysis and insights applied to all the workloads). Our first observation was that for each workload, the placement of instructions and data blocks dictated the retention time requirement. For example, in Fig. \ref{fig:Workload1}, \textit{way0} was a hot region for data block write activities and short block lifetimes. Hence, the 100$\mu$s retention time consumed the lowest energy for that way. \textit{Way1} and \textit{way3} frequently stored instructions or read-only tables; thus, the 100$ms$ retention time achieved minimum energy for those ways. Across the different workloads, we observed that the retention time requirements were indicative of the applications' execution behavior. These observations implied that energy savings can be achieved by provisioning the L2 cache with different retention time values to satisfy a variety of runtime retention needs. 

\vspace{-3pt}
\begin{figure*}[ht]
\begin{subfigure}[t]{.2\textwidth}
		\centering
		\includegraphics[width=\linewidth]{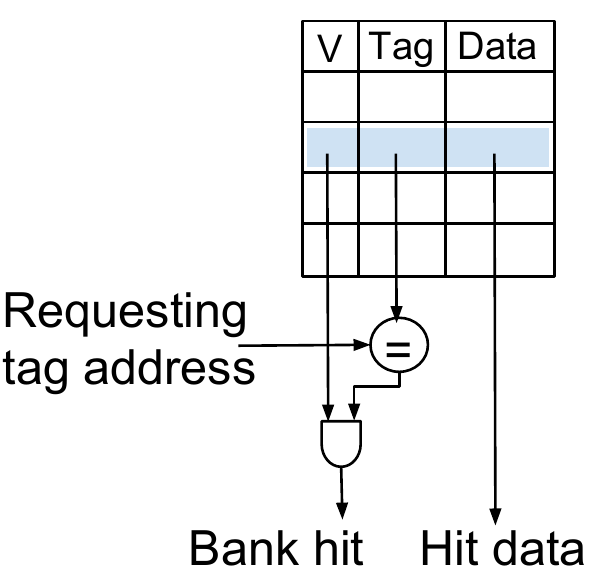}
		\caption{Detail of a cache bank}
		\label{fig:bankDetail}
\end{subfigure}
~
\begin{subfigure}[t]{.8\textwidth}
		\centering
		\includegraphics[width=\linewidth]{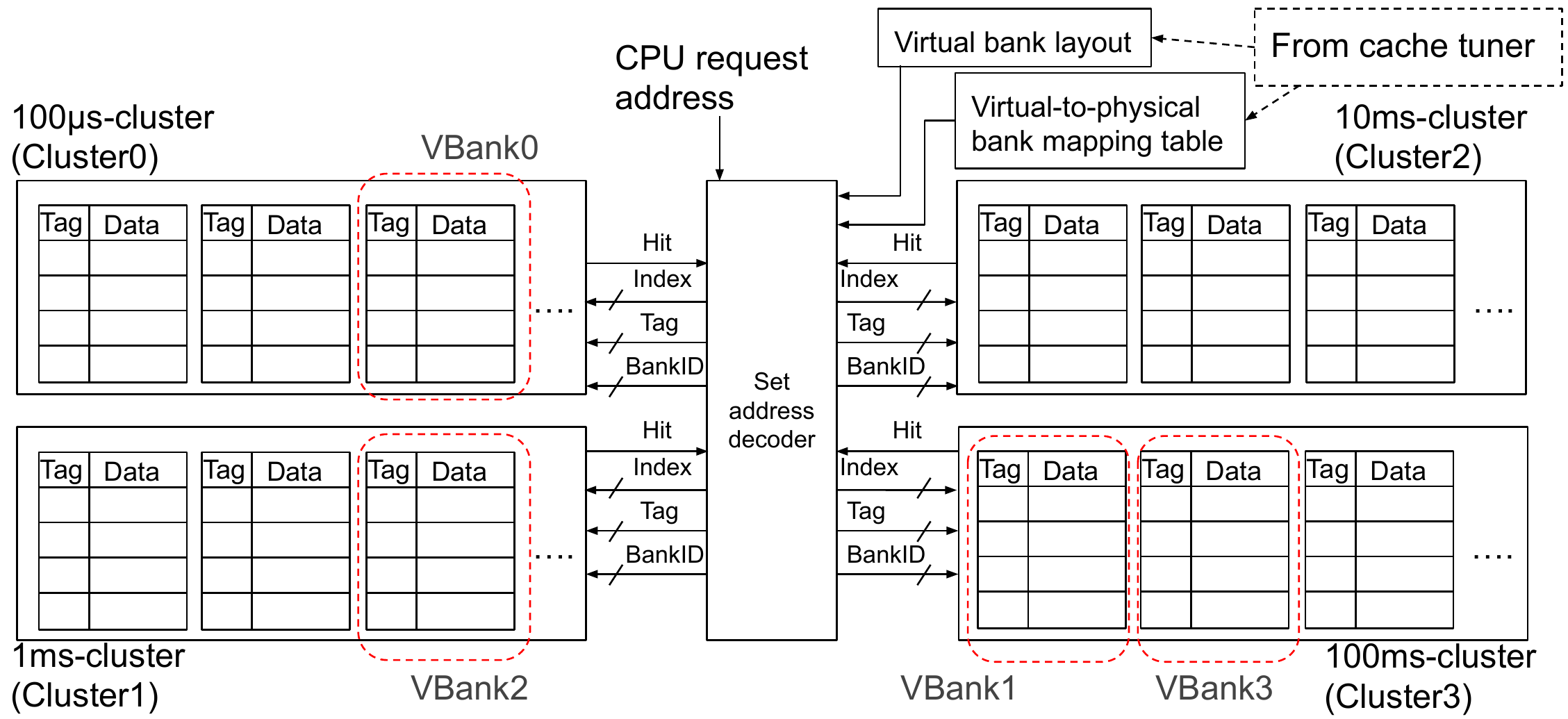}
		\caption{HALLS architecture}
		\label{fig:HALLS_arch}
\end{subfigure}
\caption{HALLS architecture}
\label{fig:HALLS_arch_all}
\end{figure*}

\begin{figure}[t]
\begin{subfigure}{.29\textwidth}

		\centering
		\includegraphics[width=\linewidth]{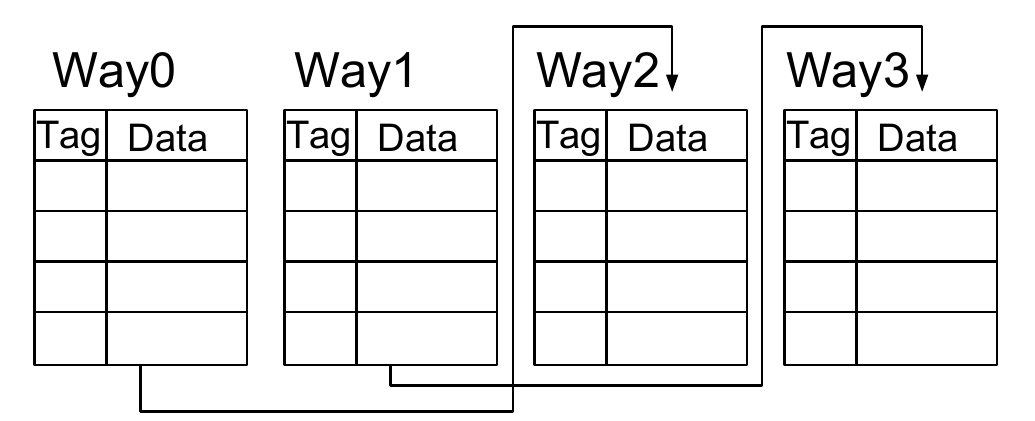}
		\caption{Configurable associativity}
		\label{fig:wayCon}
\end{subfigure}
~
\begin{subfigure}{.18\textwidth}
		\centering
		\includegraphics[width=\linewidth]{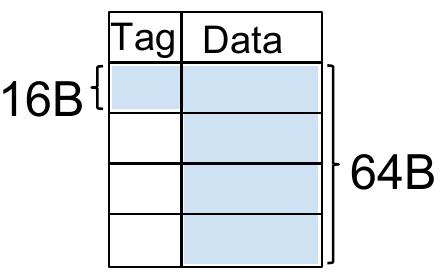}
		\caption{Configurable line size}
		\label{fig:lineCon}
\end{subfigure}
\caption{HALLS's capability of configurable associativity and line size}
\label{fig:wayLineConcatenation}
\end{figure}

\begin{figure}[ht]
		\vspace{0pt}
		\centering
		\includegraphics[width=\linewidth]{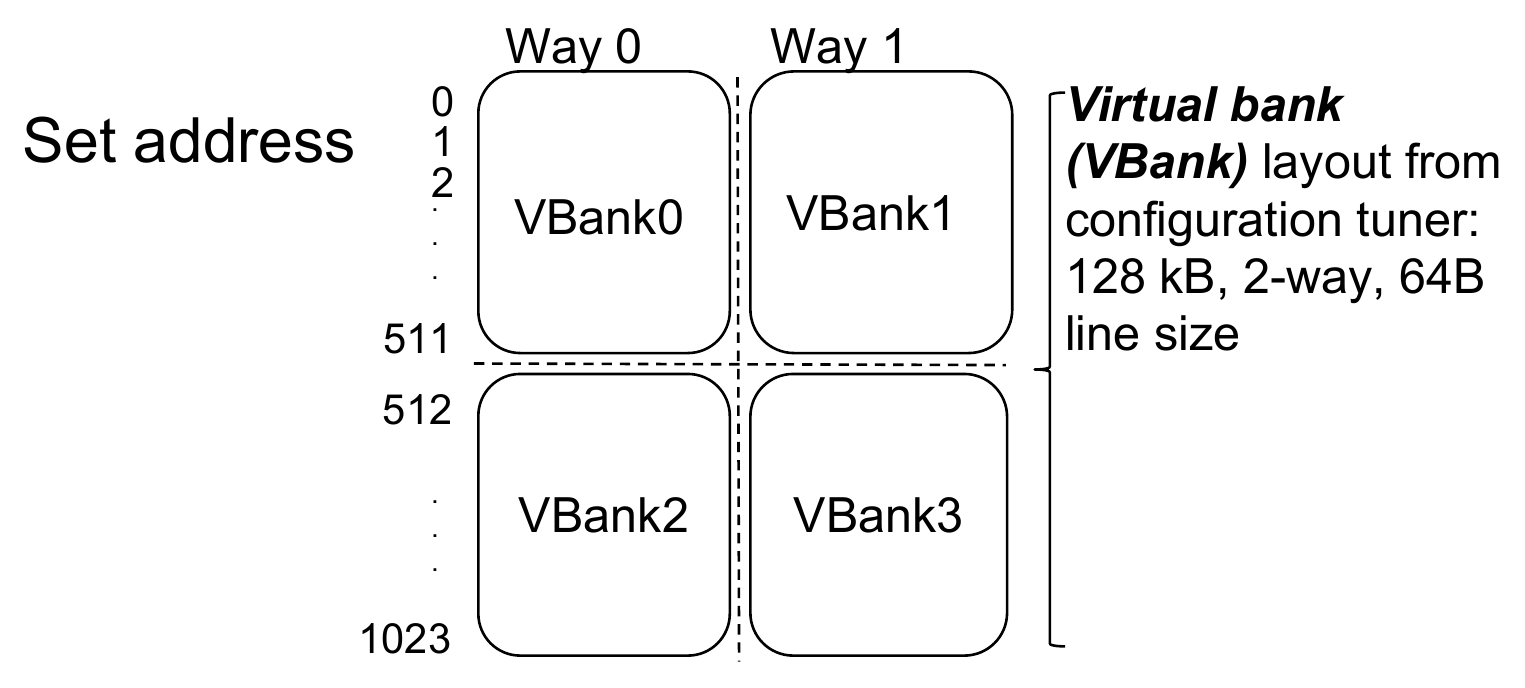}
		\vspace{-5pt}
		\caption{An example of virtual bank layout}
		\label{fig:vbank}
\end{figure}



\subsection{HALLS Architecture} \label{sec:implementation}
Fig. \ref{fig:HALLS_arch_all} illustrates the HALLS architecture. We assume a physical 1MB, 16-way STT-RAM L2 cache, since this size is found in state-of-the-art architectures \cite{a15L2}, but note that the idea can be extended to any arbitrary size L2 (or L3) cache. Since the retention time is a physical characteristic of the STT-RAM \cite{Diao07}, it cannot be easily dynamically adapted, unlike other cache configurations such as the cache size, line size, and associativity. Thus, we use a \textit{logical adaptation} of the retention time \cite{LARS}, wherein different cache banks are designed with different physical retention times. During runtime, cache blocks can be written into the cache banks that most closely match the blocks' retention time requirements.

As illustrated in Fig. \ref{fig:HALLS_arch_all}, the HALLS cache architecture is designed using 32KB banks, i.e., 32 banks in the 1MB STT-RAM cache. Runtime adaptability is achieved using similar mechanisms to SRAM-based configurable caches \cite{Zhang05}, which have been widely studied and analyzed in prior work. The cache size can be configured by shutting down cache banks, e.g., the 1MB L2 cache can be configured into a 512KB cache by shutting down 16 banks or into a 128KB cache by shutting down 28 banks. The associativity can be configured by concatenating different banks as shown in Fig. \ref{fig:wayCon}. Finally, given a physical line size of 16B, multiple lines can be fetched to logically configure the line size, from 16B to 64B, as shown in Fig. \ref{fig:lineCon}. As detailed in prior work \cite{Zhang05}, augmenting the cache for this adaptability is low-overhead and does not adversely impact the cache's critical path.

Fig. \ref{fig:bankDetail} shows details of a cache bank. Each cache bank contains a tag and a data array, along with a valid bit checker and a tag comparator. As such, each bank can operate independently, even when other banks are shut down to save power\cite{Tosi16,Chen12}. Thus, apart from enabling the adaptability proposed herein, the 32-bank structure also lends itself to higher memory bandwidth in the L2 cache. The 32 banks are organized in 8-bank clusters, with each cluster designed with a different retention time. We used a total of four retention times---100$\mu$s, 1$ms$, 10$ms$, and 100$ms$---to satisfy a range of application requirements based on empirical analysis. However, more (or different) retention times can be used depending on the executing applications. We use $ClusterID$ to indicate these four retention time clusters. Cluster0, Cluster1, Cluster2, and Cluster3 represent 100$\mu$s, 1$ms$, 10$ms$, and 100$ms$ retention time clusters, respectively.

To complete the tag look up for different configurations and different retention time clusters, we propose a modified set address decoder. As shown in Fig. \ref{fig:HALLS_arch}, the set address decoder receives the request address from CPU. Based on the cache configuration and retention time (as determined by the tuner) and the requested address, the set address decoder dispatches $Index$, $Tag$, and the corresponding $BankID$ to the appropriate retention time clusters and monitors hit bits. We define \textit{virtual bank} ($VBank$) as a specific location of cache ways and set address from the point of view of the CPU request address. Based on the outcome of cache tuning, VBank represents the mapping of requested data blocks to the physical cache banks with the appropriate retention times. Fig. \ref{fig:vbank} shows a sample layout of virtual banks, given a best cache configuration output (from the tuner) of 128 KB, 2-way, and 64B line size. VBank0 represents the cache blocks located in set address 0-511 and way 0, VBank1 represents blocks in set address 0-511 and way 1, and so on. 

The dotted boxes in Fig. \ref{fig:HALLS_arch} illustrate an example of virtual-to-physical bank mapping. When a request address has set address between 0 and 511, the set address decoder sends the index, tag, and the corresponding BankID to the 100$\mu$s-cluster (Cluster0) for VBank0 and 100ms-cluster (Cluster3) for VBank1. The BankID allows HALLS to perform bank-to-bank mapping between virtual and physical banks if multiple banks must be accessed in the same cluster. For example in Fig. \ref{fig:HALLS_arch}, 100ms-cluster (Cluster3) is able to distinguish VBank1's index and tag from VBank3's using the BankID.

\begin{figure}[t]
		\vspace{-7pt}
		\centering
		\includegraphics[width=1\linewidth]{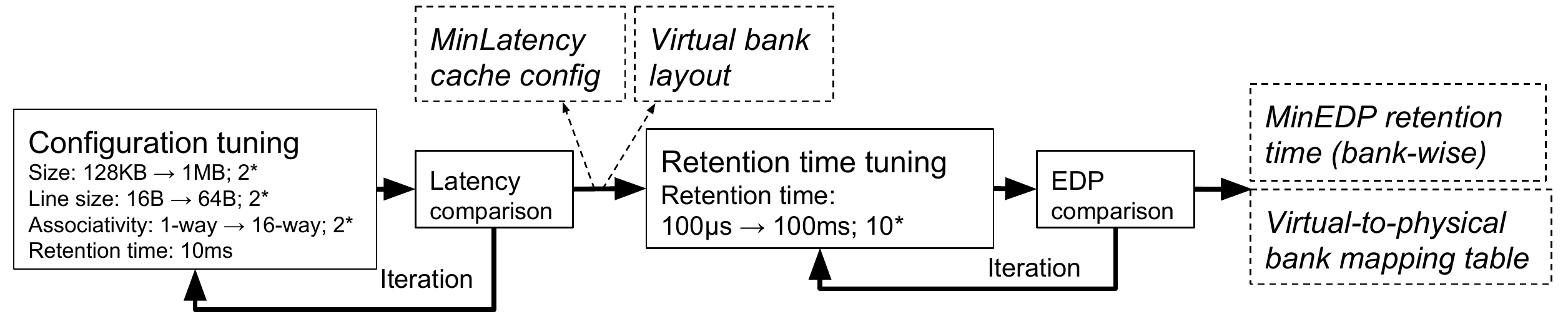}
		\vspace{-15pt}
		\caption{Block diagram of HALLS flow}
		\vspace{-7pt}
		\label{fig:HALLSdiagram}
\end{figure}

The HALLS architecture also features a low-overhead hardware \textit{tuner} that orchestrates the process of determining which cache configuration and retention time to use for an executing application. We decided to use a hardware tuner, as opposed to a software tuner, in order to make the tuner's operations non-intrusive to executing applications' behaviors (e.g., cache accesses). When a new application is run, the tuner begins sampling the application based on the tuning algorithm (Section \ref{sec:configTuner}) for a \textit{tuning interval} to determine the best (i.e. minimum energy) cache configuration. After configuration tuning, HALLS has determined the cache configuration (cache size, line size, and associativity), and then constructs a \textit{virtual bank layout} based on the determined configuration. For example, in Fig. \ref{fig:vbank}, the virtual bank layout defines a 128 KB cache with 2-way set associativity and 64B line size. Thereafter, HALLS performs the retention time tuning and establishes the mapping of the virtual banks to physical banks. Tuning details are described in Section \ref{sec:configTuner}.

Even though HALLS eliminates the need for a refresh mechanism, we still incorporate a per-block counter to keep track of the cache blocks' lifetimes \cite{Jog12}. HALLS uses the counter to detect the expiration of a cache block and evicts the block when the retention time has elapsed. Dirty blocks are first written to the main memory before eviction. We implemented the counter as a state machine, with a clock period defined as the retention time divided by $N$, where $N$ defines the granularity of block eviction. When a block is written to the cache, the counter's state advances from the initial state until it reaches the maximum state. The block is then evicted, and the counter is reset to the initial state whenever a new write operation occurs for the block. HALLS overheads are described in Section \ref{sec:results}.

\begin{algorithm}[ht]
\SetDataSty{small}
\SetKwData{MinLatency}{MinLatency}
\SetKwData{CurLatency}{CurLatency}
\SetKwData{CurConfig}{CurConfig}
\SetKwData{CurConfigp}{CurConfig.$p$}
\SetKwData{CurConfigSize}{CurConfig.size}
\SetKwData{CurConfigLineSize}{CurConfig.linesize}
\SetKwData{CurConfigWays}{CurConfig.ways}
\SetKwData{Config}{Config}
\SetKwFunction{samplingLatency}{samplingLatency}
\KwData{Size  $S=\{s_{min},...,s_{max}\}$}
\KwData{Line size  $L=\{l_{min},...,l_{max}\}$}
\KwData{Associativity  $A=\{a_{min},...,a_{max}\}$}
\KwResult{Best cache size, line size, associativity}
\BlankLine
\emph{\CurConfigSize $\leftarrow s_{max}$}\; \label{line:defaultBegin}
\emph{\CurConfigLineSize $\leftarrow l_{max}$}\;
\emph{\CurConfigWays $\leftarrow a_{max}$}\; \label{line:defaultEnd}
\emph{\Config $\leftarrow$ \CurConfig}\;
\emph{\MinLatency $\leftarrow Latency_{max}$}\;
\ForEach{$p \in [size, linesize, ways]$}{\label{line:foreachloop_begin}
\For{$(p_{i} = p_{max};  p_{i} >= p_{min};  p_{i} =p_{i} / 2)$}{ \label{line:forloop_begin}
    \emph{\CurConfigp $\leftarrow p_{i}$}\;
    \emph{\CurLatency $\leftarrow$ \samplingLatency{\CurConfig}}\;
    \If{\CurLatency $<$ \MinLatency}{ 
        \MinLatency $\leftarrow$ \CurLatency\;
        \Config $\leftarrow$ \CurConfig\;
    }
    \Else{  \label{line:else_begin}
        break\;    
    }  \label{line:else_end}
}\label{line:forloop_end}
}\label{line:foreachloop_end}
\Return{\Config,\MinLatency}\;
\caption{HALLS configuration tuning algorithm}
\label{algo:config}
\end{algorithm}

\begin{algorithm*}[ht]
\SetDataSty{small}
\SetKwData{MinEDPCluster}{MinEDPCluster}
\SetKwData{AllPhysicalBanksEDP}{AllPhysicalBanksEDP}
\SetKwData{OutputTuningSet}{OutputTuningSet}
\SetKwData{BankID}{BankID}
\SetKwData{ClusterID}{ClusterID}
\SetKwData{BankEDP}{BankEDP}
\SetKwData{MappingTable}{MappingTable}
\SetKwFunction{samplingEDP}{samplingEDP}
\SetKwFunction{findPhysicalBank}{findPhysicalBank}
\SetKwFunction{findPhysicalBankEDP}{findPhysicalBankEDP}
\SetKwFunction{popAvailablePhysicalBank}{popAvailablePhysicalBank}
\SetKwFunction{findMin}{findMin}
\KwData{Retention time tuning sets  $T=\{Set 0, Set 1, Set 2, Set 3\}$}
\KwData{Requested virtual banks $V=\{VBank0, VBank1, ..., VBank31(as\;the\;highest)\}$} \label{line:requestVBank}
\KwResult{Virtual-to-physical banks mapping table}
\BlankLine

\ForEach{$t \in T$}{
  \emph{\AllPhysicalBanksEDP $\leftarrow$ \samplingEDP{$t$}}\; \label{line:sampleEDP}
  \ForEach{$v \in V$}{
  \ClusterID,\BankID $\leftarrow$ \findPhysicalBank{$t,v$}\; \label{line:storeEDPbegin}
  \BankEDP $\leftarrow$ \findPhysicalBankEDP{$\ClusterID,\BankID,\AllPhysicalBanksEDP$}\; \label{line:storeEDPMid}
  $v$.EDP[\ClusterID] $\leftarrow$ \BankEDP\; \label{line:storeEDPEnd}
  }
}
\ForEach{$v \in V$}{
  \emph{\MinEDPCluster $\leftarrow$ \findMin{$v$.EDP}}\; \label{line:availbegin}
  \emph{\ClusterID,\BankID $\leftarrow$ \popAvailablePhysicalBank{\MinEDPCluster}}\;\label{line:availmid}
  \MappingTable.push($v,\ClusterID,\BankID$)\; \label{line:availend}

  \label{line:optimalCond2}
}

\Return{\MappingTable}\;

\caption{HALLS retention time tuning algorithm}
\label{algo:retention}
\end{algorithm*}

\subsection{Determining the best configuration} \label{sec:configTuner}
Fig. \ref{fig:HALLSdiagram} depicts the high level flow of HALLS. HALLS first determines the best cache configuration using the configuration tuning algorithm (Algorithm \ref{algo:config}). This algorithm yields the best cache configuration and the virtual bank layout. We used the latency as the algorithm's objective function based on our observation that tuning the configurations for latency improved the energy consumption, whereas tuning for energy substantially degraded the latency. We tested the algorithm using energy as the objective function and found that much smaller cache sizes were favored, thereby reducing the dynamic and leakage power. However, the resulting caches were severely under-provisioned for the applications' access requirements. On the other hand, tuning for latency allowed HALLS to tradeoff achieving optimal energy savings for reduced latency degradation.  

After determining the best cache configuration, HALLS determines the best retention time using the retention time tuning algorithm (Algorithm \ref{algo:retention}). The retention time tuning algorithm uses the energy delay product (EDP) as the objective function in order to achieve energy savings without substantial latency degradation, and outputs the virtual-to-physical bank mapping that satisfies the executing application's requirements. In this work, we used application-based tuning, but plan to explore phase-based tuning in future work. 

\subsubsection{Configuration tuning algorithm}

Algorithm \ref{algo:config} depicts the HALLS configuration tuning algorithm. The inputs to the algorithm are the cache design space, and the outputs are the best cache size, line size, and associativity. When a new application is executed, the algorithm defaults to the maximum configuration (i.e., 1MB size, 16-way set associative, and 64B line size in our HALLS architecture)  (lines \ref{line:defaultBegin}-\ref{line:defaultEnd}). HALLS then runs the application using each configuration for one tuning interval---we assumed an interval of 10M instructions in our experiments---while iterating through the configurations in descending order of the cache parameters' energy impact (cache size, followed by line size, followed by associativity). For each parameter, the configurations are explored as long as reducing the parameter value also reduces the latency (lines \ref{line:foreachloop_begin}-\ref{line:foreachloop_end}). The algorithm stops tuning if a configuration change increases the latency as compared to the current minimum latency (lines \ref{line:else_begin}-\ref{line:else_end}). The algorithm only needs to store the minimum-latency configuration, with which the current configuration's latency is compared. 

\subsubsection{Retention time tuning algorithm} \label{sec:algorithm}
To minimize implementation overhead for dynamically determining the best retention time (Fig. \ref{fig:HALLSdiagram}), we use a simple algorithm that samples all four retention times for a tuning interval. After each sampling period, HALLS collects the cache statistics from hardware performance counters and combines the statistics (e.g., read requests, write requests, writebacks, etc.) with predefined STT-RAM cache access parameters to estimate the energy consumption. 

Since the HALLS cache features retention time clusters featuring different retention times in each cluster (8 banks/cluster in our case), it is not possible to set a single uniform retention time for the cache during tuning. Thus, we defined \textit{retention time tuning sets} comprising of different retention time settings to guide the tuning process. The retention time tuning sets represent a mapping of \textit{virtual banks} (i.e., bank-sized chunks of data blocks) to physical banks to enable sampling of the applications' data with the different retention times.

Algorithm \ref{algo:retention} depicts the HALLS retention time tuning algorithm, which takes as input the retention time sets and the virtual banks, and outputs the virtual-to-physical bank mapping table. For clarity, we illustrate the retention time tuning process using Table \ref{tab:settings}, which shows the retention time tuning for the four virtual banks depicted in Fig. \ref{fig:HALLS_arch}. Assume that the best cache configuration---determined by the cache configuration tuning algorithm---is 128KB (i.e., requiring four banks), 2-way associativity, and 64B line size. The first iteration starts from \textit{Set 0}, where \textit{VBank 0,1,2,3} are allocated to Cluster0, Cluster1, Cluster2, and Cluster3, respectively. HALLS then samples the cache statistics for a tuning interval and collects the execution statistics (line \ref{line:sampleEDP}). 

Thereafter, HALLS shifts the clusters by one as shown in Table \ref{tab:settings} and collects statistics after every iteration. In every iteration, HALLS stores bank-wise EDP of physical banks into the VBank object. For each VBank, \texttt{findPhysicalBank} (line \ref{line:storeEDPbegin}) extracts the ClusterID and the BankID of the physical bank being sampled in the current iteration. Based on the collected statistics, HALLS calculates the EDP of the physical bank and stores it in VBank's EDP array (line \ref{line:storeEDPMid}-\ref{line:storeEDPEnd}). After the last iteration (\textit{Set 3}), HALLS then selects the least EDP retention time as the best for each VBank. We note that there may be a limited availability of retention times, and the best retention time may not be available if multiple VBanks select a particular retention time (e.g., 10ms for 12 banks when there are only 8 banks with 10ms). In this case, \texttt{findMin} (line \ref{line:availbegin}) searches the VBank's EDP array, checks if the physical banks in a cluster are available, and selects the best-performing available cluster. With the available ClusterID, HALLS then returns the ClusterID and the allocated BankID (line \ref{line:availmid}-\ref{line:availend}). Although we illustrate this process using only four VBanks (one for each retention time cluster), eight VBanks per cluster can be used, when necessary, to complete the tuning for the 1MB cache within four sampling intervals.

\begin{table}[ht]
\renewcommand{\arraystretch}{1.3}
\caption{Retention time tuning example}
\label{tab:settings}
\centering
\begin{adjustbox}{max width=.5\textwidth}
\begin{tabular}{|c|c||c|c|}
    \hline
    \makecell{Retention time\\tuning set}	& \makecell{VBank-Cluster\\mapping}      &\makecell{Retention time\\tuning set}				&\makecell{VBank-Cluster\\mapping}\\
    \hline
    \hline
    \multirow{4}{*}{Set 0}	&VBank0: Cluster0   & \multirow{4}{*}{Set 2}&VBank0: Cluster2\\
                            &VBank1: Cluster1   &                       &VBank1: Cluster3\\
                            &VBank2: Cluster2   &                       &VBank2: Cluster0\\
                            &VBank3: Cluster3   &                       &VBank3: Cluster1\\
    \hline
    \multirow{4}{*}{Set 1}	&VBank0: Cluster1   &\multirow{4}{*}{Set 3} &VBank0: Cluster3\\
                            &VBank1: Cluster2   &                       &VBank1: Cluster0\\
                            &VBank2: Cluster3   &                       &VBank2: Cluster1\\
                            &VBank3: Cluster0   &                       &VBank3: Cluster2\\
    \hline
\end{tabular}
\end{adjustbox}
\end{table}

\begin{table*}[ht]
\vspace{20pt}
\renewcommand{\arraystretch}{1.3}
\caption{Cache parameters of SRAM and STT-RAM for the base cache configurations}
\vspace{-5pt}
\label{tab:retention}
\centering
\begin{tabular}{c||c|cccc}
    \hline
    \hline
    Cache Configuration				&\multicolumn{5}{c}{1MB, 64B line size, 16-way}\\
    \hline
    Memory Device				&SRAM	    &STT-RAM-100$\mu$s	&STT-RAM-1ms	&STT-RAM-10ms	&STT-RAM-100ms\\
    \hline
    Write Energy (per access) 	&0.338nJ    &0.392nJ	        &0.404nJ        &0.419nJ        &0.438nJ\\
    \hline
    Cache Hit Energy (per access)	&5.318nJ    &5.794nJ	        &5.794nJ        &5.794nJ        &5.794nJ\\
    \hline
    Leakage Power               &3234.916mW     &\multicolumn{4}{c}{2200.032mW}		\\
    \hline
    Hit Latency (cycles)    	&2	        &2 			        &2		        &2 		        &2\\
    \hline
    Write Latency (cycles)    	&2	        &3			        &4		        &6		        &7\\
    \hline
\end{tabular}
\end{table*}


\begin{table*}[ht]
\vspace{10pt}
\renewcommand{\arraystretch}{1}
\caption{Cache parameters of SRAM and STT-RAM for HALLS cache configuration results}
\vspace{-5pt}
\label{tab:config_retention}
\centering
\begin{tabular}{c|c||c|cccc}
    \hline
    \hline
    \multicolumn{2}{c||}{Memory Device}				&SRAM	    &STT-RAM-100$\mu$s	&STT-RAM-1ms	&STT-RAM-10ms	&STT-RAM-100ms\\
    \hline
    \multirow{3}{*}{128K-1W-16B}&\multicolumn{1}{c||}{Write Energy (per access)} 	&0.033nJ    &0.033nJ	        &0.037nJ        &0.041nJ        &0.047nJ\\
    \cline{2-7}
    &\multicolumn{1}{c||}{Cache Hit Energy (per access)}	&0.035nJ    &\multicolumn{4}{c}{0.028nJ}\\
    \cline{2-7}
    &\multicolumn{1}{c||}{Leakage Power}               &277.744mW     &141.139mW    &141.282mW  &141.425mW  &141.568mW		\\
    \hline
    \multirow{3}{*}{128K-1W-32B}&\multicolumn{1}{c||}{Write Energy (per access)} 	&0.059nJ    &0.059nJ	        &0.066nJ        &0.074nJ        &0.084nJ\\
    \cline{2-7}
    &\multicolumn{1}{c||}{Cache Hit Energy (per access)}	&0.061nJ    &\multicolumn{4}{c}{0.051nJ}\\
    \cline{2-7}
    &\multicolumn{1}{c||}{Leakage Power}               &288.864mW     &186.218mW    &186.49mW  &186.761mW  &187.033mW		\\
    \hline
    \multirow{3}{*}{128K-2W-32B}&\multicolumn{1}{c||}{Write Energy (per access)} 	&0.058nJ    &0.056nJ	        &0.062nJ        &0.07nJ        &0.08nJ\\
    \cline{2-7}
    &\multicolumn{1}{c||}{Cache Hit Energy (per access)}	&0.117nJ   &\multicolumn{4}{c}{0.092nJ}\\
    \cline{2-7}
    &\multicolumn{1}{c||}{Leakage Power}               &346.743mW     &\multicolumn{4}{c}{185.298mW}		\\
    \hline
    \multirow{3}{*}{128K-1W-64B}&\multicolumn{1}{c||}{Write Energy (per access)} 	&0.112nJ    &0.108nJ	        &0.12nJ        &0.135nJ        &0.153nJ\\
    \cline{2-7}
    &\multicolumn{1}{c||}{Cache Hit Energy (per access)}	&0.113nJ    &\multicolumn{4}{c}{0.09nJ}\\
    \cline{2-7}
    &\multicolumn{1}{c||}{Leakage Power}               &325.697mW     &\multicolumn{4}{c}{196.05mW}		\\
    \hline
    \multirow{3}{*}{128K-4W-64B}&\multicolumn{1}{c||}{Write Energy (per access)} 	&0.130nJ    &0.150nJ	        &0.162nJ        &0.177nJ        &0.196nJ\\
    \cline{2-7}
    &\multicolumn{1}{c||}{Cache Hit Energy (per access)}	&0.519nJ    &0.519nJ	        &0.519nJ        &0.519nJ        &0.520nJ\\
    \cline{2-7}
    &\multicolumn{1}{c||}{Leakage Power}               &507.852mW     &\multicolumn{4}{c}{363.607mW}		\\
    \hline
    \multirow{3}{*}{256K-8W-64B}&\multicolumn{1}{c||}{Write Energy (per access)} 	&0.193nJ    &0.212nJ	        &0.224nJ        &0.24nJ        &0.258nJ\\
    \cline{2-7}
    &\multicolumn{1}{c||}{Cache Hit Energy (per access)}	&1.526nJ    &\multicolumn{4}{c}{1.532nJ}		\\
    \cline{2-7}
    &\multicolumn{1}{c||}{Leakage Power}               &1181.176mW     &\multicolumn{4}{c}{858.677mW}		\\
    \hline
    \multirow{3}{*}{512K-16W-64B}&\multicolumn{1}{c||}{Write Energy (per access)} 	&0.309nJ    &0.375nJ	        &0.351nJ        &0.367nJ        &0.385nJ\\
    \cline{2-7}
    &\multicolumn{1}{c||}{Cache Hit Energy (per access)}	&4.871nJ    &5.577nJ	    &4.953nJ        &4.953nJ        &4.953nJ\\
    \cline{2-7}
    &\multicolumn{1}{c||}{Leakage Power}               &2268.544mW     &1880.816mW     &1566.491mW     &1566.491mW     &1566.491mW	\\
    \hline
    \multirow{3}{*}{1M-1W-32B}&\multicolumn{1}{c||}{Write Energy (per access)} 	&0.179nJ    &0.128nJ	        &0.135nJ        &0.143nJ        &0.153nJ\\
    \cline{2-7}
    &\multicolumn{1}{c||}{Cache Hit Energy (per access)}	&0.188nJ    &0.12nJ	        &0.121nJ        &0.121nJ        &0.122nJ\\
    \cline{2-7}
    &\multicolumn{1}{c||}{Leakage Power}               &1745.328mW     &762.778mW     &763.444mW     &764.109mW     &764.775mW		\\
    \hline
    \multirow{3}{*}{1M-1W-64B}&\multicolumn{1}{c||}{Write Energy (per access)} 	&0.335nJ    &0.244nJ	        &0.257nJ        &0.273nJ        &0.291nJ\\
    \cline{2-7}
    &\multicolumn{1}{c||}{Cache Hit Energy (per access)}	&0.344nJ    &0.228nJ	        &0.229nJ        &0.229nJ        &0.23nJ\\
    \cline{2-7}
    &\multicolumn{1}{c||}{Leakage Power}               &1866.193mW     &982.701mW     &983.986mW     &985.271mW     &986.555mW	\\
    \hline
    \multirow{3}{*}{1M-2W-64B}&\multicolumn{1}{c||}{Write Energy (per access)} 	&0.329nJ    &0.25nJ	        &0.263nJ        &0.278nJ        &0.292nJ\\
    \cline{2-7}
    &\multicolumn{1}{c||}{Cache Hit Energy (per access)}	&0.663nJ    &0.464nJ	        &0.466nJ        &0.467nJ        &0.458nJ\\
    \cline{2-7}
    &\multicolumn{1}{c||}{Leakage Power}               &2276.707mW     &1472.512mW     &1475.038mW     &1477.564mW     &1456.263mW		\\
    \hline
    \multirow{3}{*}{1M-4W-64B}&\multicolumn{1}{c||}{Write Energy (per access)} 	&0.354nJ    &0.278nJ	        &0.29nJ        &0.305nJ        &0.323nJ\\
    \cline{2-7}
    &\multicolumn{1}{c||}{Cache Hit Energy (per access)}	&1.415nJ    &\multicolumn{4}{c}{1.035nJ}\\
    \cline{2-7}
    &\multicolumn{1}{c||}{Leakage Power}               &3228.278mW     &\multicolumn{4}{c}{2767.573mW}		\\
    \hline
    \multirow{3}{*}{1M-8W-64B}&\multicolumn{1}{c||}{Write Energy (per access)} 	&0.285nJ    &0.256nJ	        &0.268nJ        &0.283nJ        &0.301nJ\\
    \cline{2-7}
    &\multicolumn{1}{c||}{Cache Hit Energy (per access)}	&2.261nJ    &\multicolumn{4}{c}{1.841nJ}\\
    \cline{2-7}
    &\multicolumn{1}{c||}{Leakage Power}               &2839.156mW     &\multicolumn{4}{c}{1432.367mW}		\\
    \hline
\end{tabular}
\end{table*}

\begin{table}[t]
\vspace{10pt}
\renewcommand{\arraystretch}{1.3}
\caption{System configurations}
\vspace{-5pt}
\label{tab:memSetup}
\centering
\begin{tabular}{|c|c|}
    \hline
    System unit                 & Experimental setup\\
    \hline
    CPU				& Modeled after ARM Cortex A15 @ 2GHz \\
    \hline
    L1 SRAM I/D-Cache				& 32KB, 64 line size, 4-way, LRU, MOESI \\
    \hline
    \multirow{7}{*}{L2 STT-RAM Cache}	& Bank size: 32KB \\
                            & Size: 128KB $\longrightarrow$ 1MB; 2* \\
                            & Line size: 16B $\longrightarrow$ 64B; 2*\\
                            & Associativity: 1-way $\longrightarrow$ 16-way\\
                            & Retention time:100$\mu$s, 100$ms$, 10$ms$, 1$ms$\\
                            & Random replacement\\
    \hline
    Main memory             & 8GB DRAM\\
    \hline
\end{tabular}
\end{table}
\section{Experimental Setup} \label{sec:experiments}
We evaluated and quantified the benefits of HALLS through extensive simulations using an in-house modified version of the GEM5 simulator \cite{gem5}. We modified GEM5\footnote{The modified GEM5 version can be found at \url{www.ece.arizona.edu/tosiron/downloads.php}} to implement both HALLS and DRS to represent prior work as described in \cite{Jog12,Sun11}. To enable a stringent comparison to our approach, we modeled DRS as a "perfect" refresh scenario, meaning that there were no extra misses caused by failed refreshes, there were no unnecessary refreshes, and there were no refresh-related latency overheads. We used the modeling technique proposed in \cite{Chun13} to estimate MTJ characteristics for different retention times. Based on the technique, we calculated write pulse, write current, and MTJ resistance value $R_{AP}$ and $R_P$. With these parameters, we used NVSim \cite{NVSim} to construct the STT-RAM cache for the different retention times. We set the technology process to 22 nm to comply with the proper memory cell size that can exhibit a retention time as low as 100$\mu$s \cite{Chun13}. 

To model modern-day resource-constrained processors, we simulated dual and quad-core systems with configurations similar to processors such as the ARM Cortex A15, as shown in Table \ref{tab:memSetup}. We used retention times from 100$\mu$s to 100ms, which we empirically determined to satisfy a range of application requirements. We note that more retention times can be used at the expense of tuning complexity. 

Table \ref{tab:retention} depicts the cache parameters for the base SRAM and STT-RAM configurations. Table \ref{tab:config_retention} depicts the leakage power and dynamic energy for different configurations to illustrate how these characteristics change with different STT-RAM retention times and with respect to SRAM. The configurations are denoted as $x$K-$y$W-$z$B, where $x, y,$ and $z$ represent the cache size (in KB or MB), associativity (in ways), and line size (in B), respectively. Note that these statistics change for the different cache configurations in the design space, but for brevity, we only show the numbers for the configurations selected by our algorithm. We also only show the energy statistics, since the write latency was constant for different retention times as shown in Table \ref{tab:retention}, and hit latency was 1 cycle across the different selected configurations. The STT-RAM leakage power values resulted from the peripheral/decode circuits and optimization for read latency in our simulations. 

We observed that in the 512K-16W-64B cache, 100$\mu$s exhibited higher leakage power and dynamic energy than other higher retention times. This observation was an artifact of NVSim that we attribute to its bank organization for that configuration \cite{NVSim}. A smaller bank size was used for that cache size; as such, additional leakage was incurred from the peripheral circuits. For comparisons with HALLS, we also modeled the SRAM using NVsim \cite{NVSim}  with 22 nm technology.

\begin{table}[t]
\renewcommand{\arraystretch}{1.3}
\caption{Experimental workloads}
\vspace{-5pt}
\label{tab:workloads}
\centering
\begin{adjustbox}{max width=.48\textwidth}
\begin{tabular}{|c||c|c|}
    \hline
    Workload	&Dual-core&Quad-core\\
    \hline
    \hline
    Workload1	&calculix-leslie3d& calculix-h264ref-bzip2-leslie3d\\
    \hline
    Workload2	&omnetpp-xalancbmk& omnetpp-sjeng-gobmk-xalancbmk\\
    \hline
    Workload3	&sjeng-hmmer& sjeng-hmmer-mcf-namd\\
    \hline
    Workload4	&gobmk-bzip2& gobmk-bzip2-leslie3d-soplex\\
    \hline
    Workload5	&soplex-h264ref& soplex-h264ref-hmmer-omnetpp\\
    \hline
    Workload6	&mcf-xalancbmk& sjeng-mcf-calculix-xalancbmk\\
    \hline
    Workload7	&gobmk-h264ref& gobmk-h264ref-calculix-mcf\\
    \hline
    Workload8	&bzip2-soplex& bzip2-soplex-namd-leslie3d\\
    \hline
    Workload9	&hmmer-omnetpp& hmmer-omnetpp-h264ref-xalancbmk\\
    \hline
    Workload10	&namd-hmmer& namd-hmmer-calculix-gobmk\\
    \hline
\end{tabular}
\end{adjustbox}
\end{table}

To represent a variety of workloads, we used twelve benchmarks from the SPEC CPU2006 benchmark suite compiled for the ARM instruction set architecture, using the reference input sets. We created ten multi-programmed workload comprising of two and four randomly selected benchmarks for the dual- and quad-core experiments, respectively---with one benchmark running on each core---ensuring that all twelve benchmarks used were represented in the workloads. The workload composition is shown in Table \ref{tab:workloads}. 
\begin{figure*}[ht]
    \begin{subfigure}[t]{.48\textwidth}
		\centering
		\includegraphics[width=0.9\columnwidth]{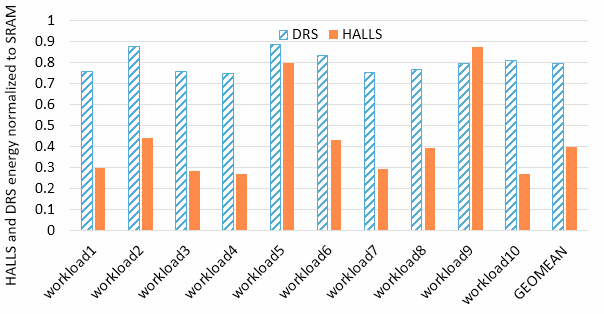}
		\caption{HALLS and DRS energy normalized to SRAM}
		\label{fig:baseOpt2coreEnergy}
    \end{subfigure}
~
    \begin{subfigure}[t]{.48\textwidth}
		\centering
		\includegraphics[width=0.98\columnwidth]{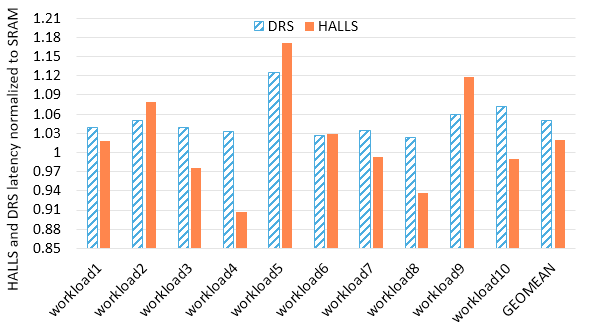}
		\caption{HALLS and DRS latency normalized to SRAM}
		\label{fig:baseOpt2coreLatency}
    \end{subfigure}
\caption{HALLS and DRS comparison in energy and latency normalized to SRAM in a dual-core system}
\label{fig:HALLS_DRS_base2}
\end{figure*}

\begin{figure*}[ht]
    \begin{subfigure}[t]{.48\textwidth}
		\centering
		\includegraphics[width=0.9\columnwidth]{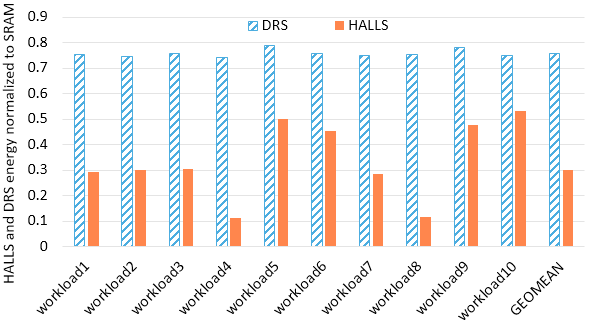}
		\caption{HALLS and DRS energy normalized to SRAM}
		\label{fig:baseOpt4coreEnergy}
    \end{subfigure}
~
    \begin{subfigure}[t]{.48\textwidth}
		\centering
		\includegraphics[width=0.98\columnwidth]{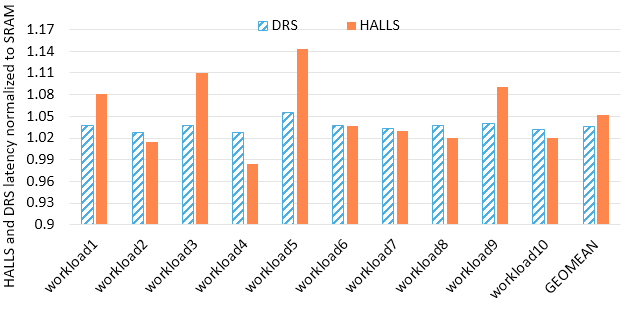}
		\caption{HALLS and DRS latency normalized to SRAM}
		\label{fig:baseOpt4coreLatency}
    \end{subfigure}
\caption{HALLS and DRS comparison in energy and latency normalized to SRAM in a quad-core system}
\label{fig:HALLS_DRS_base4}
\end{figure*}

\begin{figure*}[ht]
    \begin{subfigure}[t]{.48\textwidth}
		\centering
		\includegraphics[width=0.9\columnwidth]{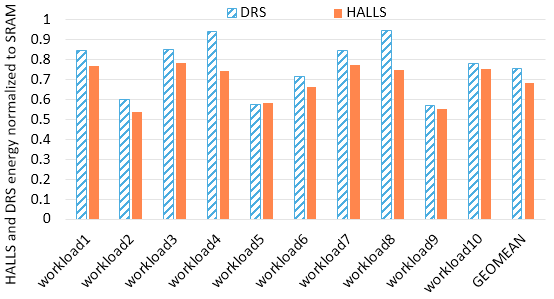}
		\caption{HALLS and DRS energy normalized to SRAM}
		\label{fig:Opt4coreEnergy}
    \end{subfigure}
~
    \begin{subfigure}[t]{.48\textwidth}
		\centering
		\includegraphics[width=0.98\columnwidth]{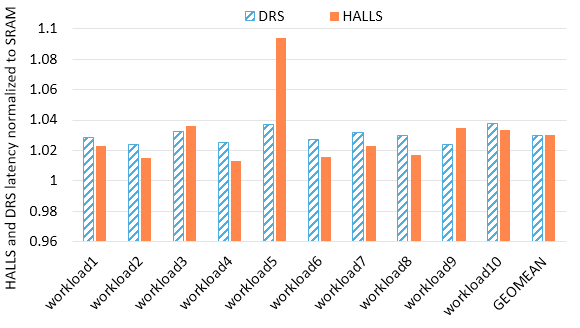}
		\caption{HALLS and DRS latency normalized to SRAM}
		\label{fig:Opt4coreLatency}
    \end{subfigure}
\caption{Energy and latency comparison of HALLS, DRS, and SRAM with adaptable cache configurations in the quad-core system. To illustrate the benefits of retention time adaptability, we assume adaptable cache configurations for all three techniques, but HALLS features adaptable retention time in addition.}
\label{fig:HALLS_DRS_opt4}
\end{figure*}

\section{Results and Comparison to Prior Work}\label{sec:results}
In both dual-core and quad-core scenarios, we evaluated HALLS's effectiveness by analyzing the L2 cache's energy as achieved by a HALLS cache compared to the SRAM and DRS with the base configuration shown in Table \ref{tab:retention}. We used a retention time of 10$ms$ for DRS, since it was considered the average best in prior work \cite{Jog12}. Prior work has shown that an SRAM LLC can consume up to 24\% of a processor's power \cite{mittal14}; thus, replacing SRAM with STT-RAM can substantially improve the energy efficiency, especially in resource-constrained systems. In general, due to a substantial reduction in leakage power, both HALLS and DRS significantly reduced the total energy as compared to the SRAM. Apart from adapting the retention time to applications' runtime needs, HALLS also provides the important advantage of enabling a right-provisioned cache for executing applications, thereby achieving substantial energy savings as compared to DRS. 

\subsection{Energy and Latency Analysis}
\subsubsection{Dual-core system}\label{sec:2core_energy_and_latency} Fig. \ref{fig:baseOpt2coreEnergy} and \ref{fig:baseOpt2coreLatency} depict the energy and latency of HALLS and DRS normalized to the SRAM in a dual-core system. On average across all the workloads, HALLS reduced the energy by 60.53\% and 50.28\% compared to SRAM and DRS, respectively. We observed substantial energy reductions specifically for $workload3$, $workload4$ and $workload10$, which contributed over 70\% and 65\% energy savings from SRAM and DRS, respectively. These workloads include $gobmk$, $namd$, $hmmer$, $bzip2$, and $sjeng$, which are compute-intensive benchmarks \cite{Phansalkar07} that exhibited short block lifetimes. As such, HALLS allocated data blocks to clusters with smaller retention time (i.e. 100$\mu$s and 1ms) and achieved smaller latency and dynamic energy.

Conversely, HALLS performed worst in energy savings for $workload5$ and $workload9$. The energy savings for $worklod5$ were 20.13\% and 9.65\% compared to SRAM and DRS, respectively. Similarly HALLS decreased the energy for $workload9$ by 12.82\% compared to SRAM and \textit{increased} by 9.59\% compared to DRS. We attribute these reduced energy savings to the fact that $workload5$ and $workload9$ include write-intensive benchmarks such as $soplex$, $h264ref$, $hmmer$, and $omnetpp$ \cite{Jog12,DASCA}. These benchmarks increased the dynamic energy due to the write operations, and also incurred additional leakage energy due to the long write latencies. 

Fig. \ref{fig:baseOpt2coreLatency} compares the latency achieved by HALLS with DRS and SRAM. On average across all the workloads, HALLS \textit{increased} the latency by 1.90\% compared to SRAM, but decreased the latency by 2.94\% compared to DRS. HALLS decreased the latency by up to 9.23\% and 12.09\% compared to SRAM and DRS for $workload4$, which contains $gobmk$ and $bzip2$, both of which are read-intensive applications and exhibit short block lifetimes. Due to the STT-RAM's short hit latency and HALLS's ability to adapt the configurations to the executing workloads' requirements, HALLS reduced the latency for five of ten workloads compared to SRAM. However, for $workload5$---a write-intensive workload---HALLS increased the latency by up to 17.10\% and 4.07\% compared to SRAM and DRS, representing a substantial latency tradeoff in favor of the aforementioned energy savings.

\subsubsection{Quad-core system}\label{sec:4core_energy_and_latency} Fig. \ref{fig:baseOpt4coreEnergy} and \ref{fig:baseOpt4coreLatency} depict the energy and latency of both HALLS and DRS normalized to SRAM in a quad-core system. On average across the workloads, Fig. \ref{fig:baseOpt4coreEnergy} shows that HALLS reduced the average energy by 60.57\% and 70.12\%, as compared to DRS and SRAM, respectively. Compared to DRS, energy savings were over 85\% for $workload4$ and $workload8$. This improvement was possible due to the short block lifetimes of the benchmarks featured in this workload. Both of these workloads featured $bzip2$ and $leslie3d$ (Table \ref{tab:workloads}), both of which featured blocks that exhibited short lifetimes. Unlike most other benchmarks, both $bzip2$ and $leslie3d$ also had low increase in miss rates for low retention times compared to the higher retention times. As such, HALLS was able to use a short retention time, while also adapting the cache configurations to the benchmarks' requirements. In most cases, HALLS's adaptability reduced the energy by more than 50\%. These results illustrate HALLS's ability to adapt STT-RAM configurations to the variety of execution requirements exhibited by applications in multicore systems. 

We note that HALLS's substantial energy reduction was at the expense of some latency overhead. Fig. \ref{fig:baseOpt4coreLatency} shows that HALLS increased the latency by 5.16\% and 1.47\%, as compared to SRAM and DRS, respectively. Latency overheads were up to 14.25\% for $workload5$ as compared to SRAM. These latency overheads occurred because HALLS resulted in additional misses for several data blocks whose lifetimes exceeded the available retention time. We also observed that another important factor that caused the latency overhead was the workloads' reactions to the STT-RAM's long write latency characteristic. We observed that the most latency degradation occurred with workloads that featured write-intensive applications. For instance, $workload5$ contained $soplex$, $h264ref$, and $omnetpp$, which are characterized as write-intensive benchmarks \cite{Jog12,DASCA}. Similarly, $workload3$ also featured three write-intensive benchmarks: $sjeng$, $hmmer$, and $mcf$. 

Compared to DRS, HALLS's latency overheads were down to 6.97\% and 8.22\% for $workload3$ and $workload5$, respectively. We reiterate that considering our target of resource-constrained systems, the average 1.47\% latency overhead can be considered an acceptable tradeoff for the substantial 60.57\% energy savings compared to DRS. 

\subsubsection{Summary of HALLS's latency behaviors}\label{sec:latency_summary}
We observed that workloads exhibited similar latency behavior in HALLS for both dual-core and quad-core systems. We observed a strong correlation between HALLS's performance (compared to SRAM and DRS) and the write intensities and cache block lifetimes of the benchmarks featured in executing workloads. To illustrate these observations, Table \ref{tab:latency_summary} depicts how HALLS's average latency compared with SRAM and DRS for workloads featuring read vs write intensive benchmarks, and short vs. long block lifetimes. We defined a benchmark as having short block lifetimes if the benchmark's data blocks were not required in the cache (i.e., time between successive references) beyond 1ms on average, while benchmarks with long block lifetimes were required in the cache for more than 1ms. 

While SRAM outperformed HALLS for write-intensive workloads, SRAM's superiority over HALLS for latency was less visible for workloads that had short block lifetimes. This behavior is exemplified in the dual-core system by $workload5$ and $workload9$ (Table \ref{tab:workloads}). All four benchmarks in these two workloads are write-intensive benchmarks. However, $hmmer$ (in $workload9$) exhibits shorter cache block lifetimes than $omnetpp, soplex,$ and $h264ref$. Thus, compared to SRAM, HALLS increased the latency for $workload9$ by a smaller amount (11.83\%) than $workload5$ (17.10\%). 

We also observed this behavior between the dual- and quad-core systems. For $workload5$, for example, HALLS increased the latency compared to SRAM by 17.10\% and 14.25\% in the dual- and quad-core systems, respectively. In the dual-core system, $workload5$ comprises of two write-intensive benchmarks ($soplex$ and $h264ref$) with long block lifetimes. The introduction of $hmmer$, which has short block lifetimes, to the mix for the quad-core system caused a latency reduction, even though the other three benchmarks had longer cache block lifetimes. We also observed that HALLS performed best with respect to latency for read-intensive benchmarks with shorter cache block lifetime. For instance, workloads featuring $bzip2$, $gobmk$, or $xalancbmk$ ($workload2,4,6,7,8$) exhibited small latency overheads in both the dual-core (1.28\% on average) and quad-core systems (1.66\% on average) (Fig. \ref{fig:baseOpt2coreLatency} and \ref{fig:baseOpt4coreLatency}). As shown in Table \ref{tab:retention}, in the same cache configuration, write latency in STT-RAM is generally higher than SRAM, and the latency grows as the retention time increases. As such, read-intensive benchmarks that issue fewer write requests would suffer less latency overheads as compared to SRAM. If the benchmark also exhibits shorter cache block lifetime, HALLS can adapt to a shorter retention time requirement without substantial overheads from cache misses, thereby taking advantage of shorter write latency per access.

\begin{table}[b]
\renewcommand{\arraystretch}{1.3}
\caption{Summary of latency behaviors in HALLS}
\vspace{-5pt}
\label{tab:latency_summary}
\centering
\begin{tabular}{|c|c|c|c|c|}
    \hline
    Benchmarks	& Intensity & \makecell{Cache block\\lifetime} & \makecell{HALLS\\vs SRAM} & \makecell{HALLS\\vs DRS}\\
    \hline
    \hline
    \makecell{hmmer,leslie3d,\\sjeng}	&write  & short   & +4.29\%   &+0.45\%\\
    \hline
    \makecell{soplex,h264ref,\\mcf,omnetpp}	&write  & long    &+13.03\%   &+5.67\%\\
    \hline
    \makecell{bzip2,xalancbmk,\\gobmk}	&read   & short   &-5.44\%   &-8.18\%\\
    \hline
    calculix,namd	&read   & long    &+2.68\%   &-1.67\%\\
    \hline
\end{tabular}
\end{table}

\subsubsection{Benefits of retention time adaptability} To explore the benefits of exclusively adapting the retention time using our approach in a shared L2 cache, we also implemented and analyzed DRS and SRAM with adaptable cache configurations as determined by the HALLS configuration tuning algorithm. That is, DRS's retention time was kept constant, while its (and SRAM's) cache configurations were adapted to the different applications' requirements similar to HALLS. 

Fig. \ref{fig:Opt4coreEnergy} depicts the energy and latency comparison for HALLS and DRS, normalized to SRAM, given adaptable cache configurations for all three techniques. On average across all the workloads, HALLS reduced the energy by 31.83\% and 9.34\% as compared to SRAM and DRS, respectively. We observed energy savings (compared to DRS) as high as 21.01\% and 20.84\% for $workload4$ and $workload8$, respectively. As described earlier, we attribute these energy savings to the benchmarks' short block lifetimes. HALLS took advantage of the energy benefits of smaller retention times that more closely match the applications' needs.

With the same configuration across HALLS, SRAM, and DRS, HALLS (with variable retention times) incurred smaller latency overhead than when compared to static SRAM and DRS configurations. As shown in Fig. \ref{fig:Opt4coreLatency}, on average, HALLS increased the latency by 3.02\% as compared to SRAM, and exhibited nearly the same latency on average (with a marginal 0.03\% improvement) as compared to DRS. Compared to DRS, HALLS marginally \textit{decreased} the latency for seven out of ten workloads with the highest reduction at 1.28\% for $workload8$. The highest latency overhead was 5.48\% for $workload5$. Apart from $workload5$, other workloads' latency overheads were below 1\%.

\subsection{HALLS Overheads:}
 HALLS overheads result from: 1) hardware overhead: the tuner---including the datapath for energy estimation---and the counter (Section \ref{sec:implementation}); 2) time overhead, including the runtime tuning overhead, which includes the time it takes to determine the best configuration, and context switching overheads when swapping cache data during reconfiguration.

We estimated the tuner overhead using Verilog and analyzed using Synopsys Design Compiler. The estimated area overhead was 0.0145 $mm^2$, and the dynamic and leakage power were 28.04 mW and 422.68 $\mu$W, respectively. Compared to an ARM Cortex-A15 processor \cite{CortexA15area}, the tuner's overhead is negligible (approximately 0.095\%). The counter required 4 bits per 64B block, resulting in a 0.78\% overhead. We assume the counter is stored in the STT-RAM along with other meta-data (e.g., tags), in order to further reduce area and power overheads. 

We also evaluated the number of tuning intervals required to determine the best configuration for the different applications. Overall, the highest number of intervals required was seven to determine the best cache configuration. Retention time tuning took a constant of four tuning intervals for all applications, since all the retention times were sampled. Given our tuning interval of 10M instructions, the tuning overheads amortize rapidly over the rest of the application's execution. In the worse case, context switching incurred a latency of 114688 cycles and an energy overhead of 14.844 $\mu$J.

\section{Conclusions and Future Research}
In this paper, we propose a \textit{highly adaptable last level STT-RAM cache (HALLS)} as a viable option for mitigating the overheads of implementing the STT-RAM in last level caches (LLC). HALLS allows the LLC's configurations to be dynamically adapted to executing applications' cache configuration and retention time requirements. We designed HALLS as a 1MB L2 cache organized as 32 physical banks. The 32 banks are organized in 8-bank clusters, with each cluster featuring a different retention time. During runtime, data blocks are placed in the physical banks that best suits the applications' retention time requirements. Furthermore, the cache configuration can be adapted to suit executing applications' needs. Experiments reveal that in a quad-core system, HALLS reduced the average energy by 60.57\% compared to prior work, while introducing 1.47\% of latency overhead. For future research, we plan to explore the design space of retention times for multithreaded applications and also explore techniques for reducing runtime overheads using history-based prediction of the best cache configurations and retention times.

%
	\bibliographystyle{IEEEtran}
	\bibliography{refs}
\vspace{150pt}
\begin{IEEEbiography}[{\includegraphics[width=1in,height=1.25in,clip,keepaspectratio]{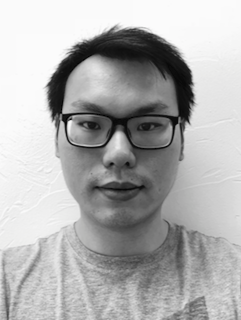}}]{Kyle Kuan} (M'16) is a Ph.D. student in the Department of Electrical and Computer Engineering at the University of Arizona. He received his M.S. in Electrical Engineering from National Taiwan University in 2008 and B.S. in Mechanical Engineering from National Chiao Tung University in 2006. His research interests include cache design for energy efficient systems, non-volatile memories, and right-provisioned micro architectures for IoT devices.
\end{IEEEbiography}
\vspace{-430pt}
\begin{IEEEbiography}[{\includegraphics[width=1in,height=1.25in,clip,keepaspectratio]{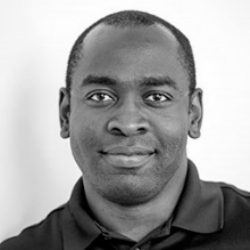}}]{Tosiron Adegbija} (M'11) received his M.S and Ph.D in Electrical and Computer Engineering from the University of Florida in 2011 and 2015, respectively and his B.Eng in Electrical Engineering from the University of Ilorin, Nigeria in 2005. 

He is currently an Assistant Professor of Electrical and Computer Engineering at the University of Arizona, USA. His research interests are in computer architecture, with emphasis on adaptable computing, low-power embedded systems design and optimization methodologies, and microprocessor optimizations for the Internet of Things (IoT). 

Dr. Adegbija was a recipient of the CAREER Award from the National Science Foundation in 2019 and the Best Paper Award at the Ph.D forum of IEEE Computer Society Annual Symposium on VLSI (ISVLSI) in 2014.
\end{IEEEbiography}
	
\end{document}